\documentclass[12pt]{article}
\usepackage[margin=1in]{geometry}
\usepackage{authblk}
\usepackage{amsmath,amsthm,amsfonts,amssymb}
\usepackage{enumitem}
\usepackage{algorithm,algpseudocode}
\usepackage{hyperref}
\usepackage{apacite}
\usepackage{bbm}
\usepackage{graphicx}
\usepackage{setspace}
\usepackage{mdframed}
\usepackage{sectsty}
\usepackage{titlesec}
\usepackage{graphicx}
\usepackage{subcaption}
\usepackage[parfill]{parskip}
\usepackage[normalem]{ulem}
\useunder{\uline}{\ul}{}

\makeatletter
\patchcmd{\@maketitle}{\LARGE \@title}{\fontsize{16}{16}\selectfont\@title}{}{}
\makeatother
\sectionfont{\fontsize{12}{12}\selectfont}
\subsectionfont{\fontsize{12}{12}\selectfont}
\begin{document}
\title{\textbf{ BayesSpeech: A Bayesian Transformer Network for Automatic Speech Recognition }}
\author{\textbf{Will Rieger}}
\affil{ 
    \textit{Master of Science in Computer Science}
    \protect \\ 
    \textit{Department of Computer Science}
    \protect \\ 
    \textit{The University of Texas at Austin}
}
    
\date{}
\maketitle
\vspace{-2em}

\begin{mdframed}[leftmargin=20pt, rightmargin=20pt]
\begin{abstract}

Recent developments using End-to-End Deep Learning models have been shown to have near 
or better performance than state of the art Recurrent Neural Networks (RNNs) on 
Automatic Speech Recognition tasks. 
These models tend to be lighter weight and require less training time than traditional RNN-based approaches. 
However, these models take frequentist approach to weight training.
In theory, network weights are drawn from a latent, intractable probability distribution. 
We introduce BayesSpeech for end-to-end Automatic Speech Recognition. 
BayesSpeech is a Bayesian Transformer Network where these intractable posteriors are learned through 
variational inference and the local reparameterization trick without recurrence.  
We show how the introduction of variance in the weights leads to faster training time and near state-of-the-art performance on LibriSpeech-960.

\end{abstract}
\end{mdframed}

\onehalfspacing

\section*{ 1. Introduction }
\vspace{-0.5em}

In the majority of neural networks, randomness is usually introduced through perturbation of the input or randomly removing nodes from the network \shortcite{dropout}.
There has been great success using these methods across a variety of domains including Automatic Speech Recognition \shortcite{spec-augment}.
Models continue to evolve. 
However and data augmentation methods rarely take large leaps in terms of the features they can help express.
Newer models are generally larger and larger and require incredible amounts of compute to properly train.
We especially see this in the field of Automatic Speech Recognition.
Newer models such as Jasper \shortcite{jasper}, the Conformer \shortcite{conformer}, LAS \shortcite{LAS}, and the Transformer \shortcite{attn-is-all-u-need}
all require training for multiple days across multiple GPUs.
Creating deeper models can certainly help attain better performance on the domain task.
But what if we approach the model differently and try to leverage their probabilistic nature?

Most Neural Network models take a frequentist approach to model training.
As you introduce non-linearities and apply gradient methods to solving these optimization problems, we become less and less likely to know we have reached a true minima.
In theory, our true weights are drawn from an intractable prior distribution. 
If we approach the problem through a Bayesian lens, we can better contextualize our model's output and weights on the input data.
Using variational inference techniques, we can design a network whose weights are drawn from a learnable, tractable posterior.
We present BayesSpeech; a Bayesian Transformer Network for End-to-End Automatic Speech recognition where feed forward layers are contextualized with probability distributions.

\section*{ 2. Background }
\vspace{-0.5em}
\section*{ 2.1 Automatic Speech Recognition }
\vspace{-0.5em}
Automatic Speech Recognition models have been evolving rapidly in recent years.
Models can either be sub-domain specific and focus on speech representation \shortcite{acoustic-modeling} \shortcite{audio-classif-dbn} \shortcite{cross-lingual-learning} \shortcite{BERT-pre-training} \shortcite{wav2vec-pre-training} \shortcite{autoreg-speech-repr} \shortcite{rectifier-for-acoustic} \shortcite{wav2vec2-self-supervised}
attention \shortcite{attn-is-all-u-need} \shortcite{attention-based-models-asr} \shortcite{cross-lingual-learning} or be end-to-end and incorporate the aforementioned components into one model and jointly train them.

\section*{ 2.1.1 Connectionist Transporal Classification }
\vspace{-0.5em}
In order to jointly train end-to-end model including alignment, and encoding/decoding the input/output sequence, Connectionist Transporal (CTC) Loss can be used to better manage the alignments \shortcite{CTC-graves}.
Alignment of the input and output sequence becomes especially challenging in speech recognition tasks as the input sequence is generally longer than the output sequence.
CTC Loss aids this process by penalizing models based on the joint probability of the current token in the sequence and all other tokens predicted.

For decoders that only output set-length sequences, we can further augment the CTC loss by utilizing traditional Cross Entropy loss on the predictions \shortcite{joint-ctc-attn}.
By enabling a joint CTC and Cross Entropy (CE) loss function we not only penalize characters based on their sequence but also on their absolute positioning in the output. 
This is covered further in Section 3.2.2.

\section*{ 2.1.2 Models for End-to-End Speech Recognition}
\vspace{-0.5em}

End-to-End Speech recognition models combine all of the individual aspects of Automatic Speech Recognition into one model that is trained jointly.
Traditional models rely on RNNs, LSTMs, and generally recurrences for defining the output sequence \shortcite{towards-e2e-unsupervised-asr}\shortcite{LAS}.
These models are cumbersome to train and parallelize and create additional operational hurdles in properly tuning.

Recently, new models involving convolutions and linear outputs have been used within Encoder-Decoder frameworks for end-to-end speech recognition tasks.
These models such as Jasper \shortcite{jasper}, Conformer \shortcite{conformer}, and Transformer \shortcite{speech-transformer} are huge models with one billion or more parameters relying on feed-forward architectures.
The three models were all trained for multiple days on multiple GPUs and required incredible compute power. 
The Speech-Transformer model \shortcite{speech-transformer} tried to address these issues by creating a thinner model to yield similar performance on the WSJ dataset.
However, compared to its larger counterparts, it did not have the same performance characteristics. 
Although it did lend hope that smaller models could be trained to compete with their larger counterparts.

\section*{ 2.2 Bayesian Methods }
\vspace{-0.5em}

Bayesian models have begun to show further promise in multiple fields such as Image Recognition \shortcite{weight-uncertainty}, attention mechanisms \shortcite{bayesian-attention-dbn} \shortcite{bayesian-attention-modules}, and auto-encoders \shortcite{auto-encoding-variational-bayes}.
In a Bayesian approach, networks weights are samples from an intractable distribution which we can estimate over training iterations through Variational Inference.

\section*{ 2.2.1 Variational Inference }
\vspace{-0.5em}

Variational Inference (VI) is the estimation of an intractable distribution through minimizing the Kullback-Lieblier divergence ($D_{KL}$) between a sample and some true distribution.
Different works have shown that these estimation methods have value when applied to a Bayesian Neural Network \shortcite{practical-vi} \shortcite{auto-encoding-variational-bayes}.
There are two different approaches to VI that largely depend on the what the true distribution is believed to be. 
If the true prior can be any distribution Monte-Carlo sampling is the only option for estimating the gradient from $D_{KL}$.
When using MC sampling, a network is sampled multiple times for the same input and the gradients are averaged together across the number of samples.

If the prior is generally assumed to be Gaussian, the KL Divergence can be explicitly calculated and the Local Reparameterization Trick \shortcite{local-reparam} can be used for finding the gradient with just one sample.
This is discussed further in Section 3.1.4.

\section*{ 2.2.2 Bayes by Backprop }
\vspace{-0.5em}
Blundell et al. introduced the Bayes by Backprop algorithm for jointly learning the intractable distribution as well as a domain problem in a Bayesian Neural Network.
They introduce the joint loss function in two parts:
\begin{enumerate}
    \item Weighting the KL Divergence of the model against the epoch
    \item Creating an Evidence Lower Bound (ELBO) on the loss function for the purpose of training
\end{enumerate}
We discuss the weighting of the KL Divergence loss term over time in Section 3.2.1.
The introduction of ELBO loss serves as the method for tuning an efficient approximator for the Maximum Likelihood given an a-posteriori inference of the parameters.
Because we are always sampling from an intractable distribution, the KL divergence term can be thought of as a regularization constant on the network.
Over time, the KL divergence impact on loss will encourage the approximate posterior to be close to the true prior.
While not readily apparent, ELBO loss is implicitly involved in the loss function described in Section 3.2.3.

\section*{ 3. Research \& Methods }
\vspace{-0.5em}

As introduced above, sampling network weights (Bayesian approach) rather than explicitly defining them (frequentist approach) has been shown to have 
increased performance and faster convergence times.
Our goal is to produce a network, leveraging Bayesian layers, to compete with state-of-the-art models and require less training time.  
BayesSpeech, is largely based on the no-recurrence model, the Transformer \shortcite{attn-is-all-u-need}. 
While we leverage the model's general architecture, we introduce modified Encoder and Decoder layers with Bayesian, Pointwise Feed Forward sub-layers.
In this section, we explore the core components of the model, the model's architecture, and a new training methodology for an ensemble loss function.

\section*{ 3.1 Core Components }
\vspace{-0.5em}
\section*{3.1.1 Attention Mechanisms }
\vspace{-0.5em}
Part of Vaswani et al.'s Transformer \shortcite{attn-is-all-u-need} network was the introduction of Scaled Dot-Product 
attention and, further, Multi-Headed Attention.
The goal of these mechanisms is to generate a temporal-rich representation of the inputs by attending 
to different positions within the input sequence.
An attention function maps a query (or set of queries) in a matrix, {$Q$}, and a set of key-value pairs in matrices, {$K, V$}, to the input sequence.

Scaled Dot-Product Attention first takes the softmax of the matrix multiplication of {$Q$}, {$K$}.
Then it matrix multiplies that value with $V$ and normalizes by {$\sqrt{d_k}$} (the dimension of the keys, {$K$}) (Equation \ref{eq:dpattn}).
The normalization by the key size is used to prevent the softmax function from suffering the vanishing gradient problem.
\begin{equation}
    Attention(Q, K, V) = Softmax(\frac{QK^T}{\sqrt{d_k}})V
    \label{eq:dpattn}
\end{equation}

Scaled Dot-Product attention only performs a single attention function at a time.
Multi-Head Attention addresses this by linearly projecting the queries, keys, and values
$h$ times to each of the input dimensions ($d_q$, $d_k$, $d_v$, respectively).
Then Scaled Dot-Product attention is applied to these newly projected inputs in order to 
attend across the $h$ different "heads" (Equation \ref{eq:mhattn}). 
Each $head_i$ is equal to $Attention(QW^Q_i, KW^K_i, VW^V_i)$.
This way the model jointly attends different input representations across the different subspaces
introduced through the projection.
\begin{equation}
    MultiHead(Q, K, V) = Concat( head_1, ..., head_h )W^O 
    \label{eq:mhattn}
\end{equation}

\section*{ 3.1.2 Sinusoidal Positional Encoding }
\vspace{-0.5em}
Because the model does not have recurrence or convolutions, in order to make use of the temporal attentions
from the Multi-Head modules, position information about the position of the tokens in the sequence must be introduced \shortcite{attn-is-all-u-need}.
Positional Encodings are used in the both the Encoder and Decoder modules to align each module's outputs and allow them to be summed.
The model leverages Sinusoidal embeddings (Equation \ref{eq:sinemb}) where the frequency corresponds to the token position ($pos$) and the dimension ($i$). 
Vaswani et al. hypothesize that using this encoding function will make it easy for the model to learn the attention weights for relative positions
and adapt for longer sequences.

\begin{equation}
    PositionalEmbedding_{(pos, i)} = \begin{cases}
        sin(pos/10000^{\frac{2^i}{d_{model}}}) & \text{if $i < \frac{d_{model}}{2}$} \\
        cos(pos/10000^{\frac{2^i}{d_{model}}}) & \text{if $i \geq \frac{d_{model}}{2}$} 
    \end{cases}
    \label{eq:sinemb}
\end{equation}

\section*{ 3.1.4 Proposal: Bayesian, Positionwise Feed Forward Layer }
\vspace{-0.5em}
\begin{figure}[!hb]
    \centering
    \includegraphics[scale=0.20]{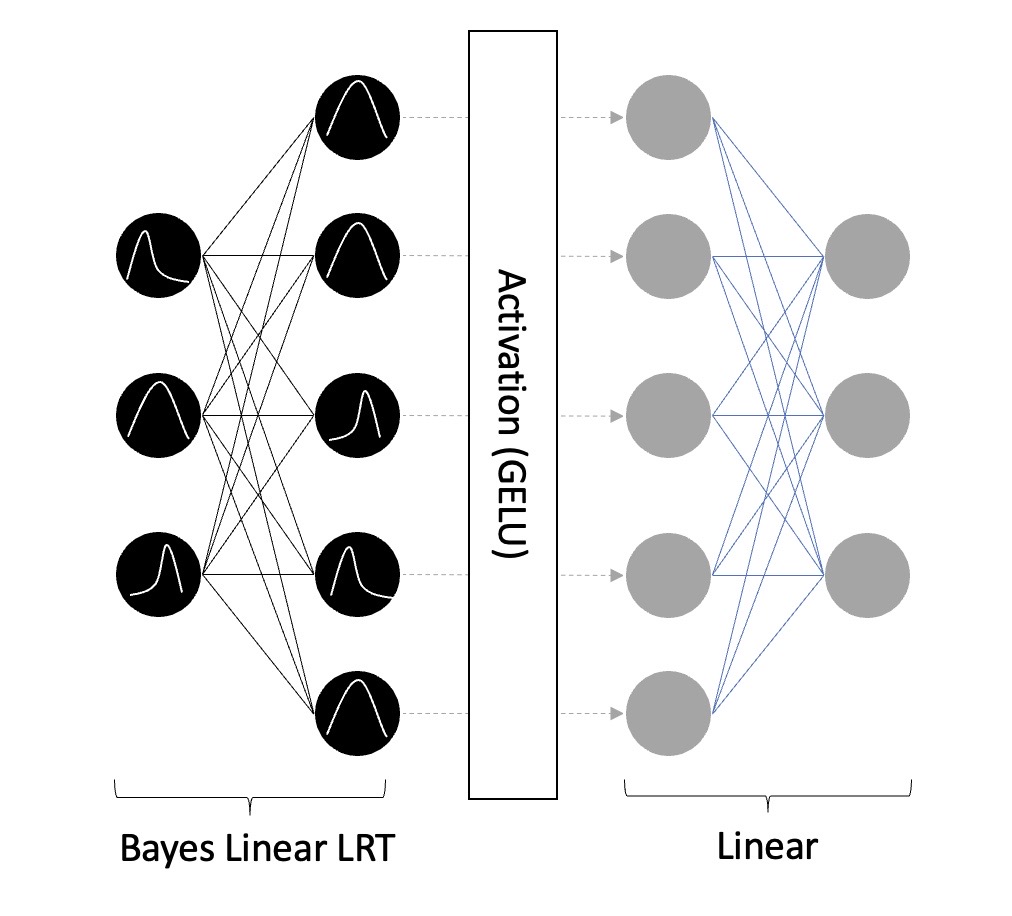}
    \caption{ Bayesian, Positionwise Feed Forward Layer Diagram}
    \label{fig:BPWFF}
\end{figure}

Our primary proposal, is the Bayesian, Positionwise Feed Forward Layer (Figure \ref{fig:BPWFF}).
Rather than use two linear layers with dropout, we substitute the input layer with a Bayesian Linear Layer suplemented with the 
Local Reparameterization Trick (Bayes Linear LRT) \shortcite{local-reparam} and remove the dropout.
Other approaches to producing Bayesian components rely on sampling the same input sequence multiple times from a Monte-Carlo process in order to
define an average gradient for modeling the intractable posterior \shortcite{weight-uncertainty}.
In addition to being computationally expensive, this can also lead to high variance in the gradients increasing training time.
The Local Reparameterization Trick addresses this by assuming that both the prior and posterior are Gaussian.
Using only a single sample, the KL divergence between the estimators can be explicitly solved for.
This new estimator is efficient (as it has less computational complexity) and reduces the variance in the gradient.

For each Bayesian Layer in Figure \ref{fig:BPWFF}, let $d_{in}$, $d_{out}$ be the input and output dimensions of the layer, respectiveley.
We define matrices $W_{\mu} \in \mathbb{R}^{d_{out}X d_{in}}$ and $W_\rho \in \mathbb{R}^{d_{out}X d_{in}}$ representing the mean and variance scalar for each weight in the network.
Similarly we have bias vectors for the output of $W_\mu$ and $W_\rho$ defined as $b_\mu \in \mathbb{R}^{d_{out}}$ and $b_\rho \in \mathbb{R}^{d_{out}}$, respectively.
Finally, at each iteration we sample a term from a standard Gaussian, $\epsilon \sim \mathcal{N}(0, 1)$.

In order to calculate the output of the layer, we define a function for explicitly calculating the KL divergence when the prior and posterior are both Gaussian (Equation \ref{eq:KLD}).
The prior is represented by $p$ and posterior represented by $q$. A sum is taken over all elements in the input matrices.
\begin{equation}
    KLD(\mu_{p}, \sigma_{p}, \mu_{q}, \sigma_{q}) = \frac{1}{2} \sum ( 2 \log( \frac{\sigma_{p}}{\sigma_{q}}) - ( 1 + (\frac{\sigma_{p}}{\sigma_{q}})^2) + (\frac{ \mu_{p} - \mu_{q}}{\sigma_{p}})^2)
    \label{eq:KLD}
\end{equation}

In the forward pass of the algorithm, we first must use our variance parameter $\rho$ for estimating our standard deviation of weight and biases (Equation \ref{eq:eststd}).
The same applies for the bias vector $b$ (e.g. we arrive at a $b_{\sigma}$ using $b_{\rho}$).
\begin{equation}
    W_{\sigma} = \log( 1 + e^{ W_{\rho} } )
    \label{eq:eststd}
\end{equation}
Next we sample from our Gaussian and introduce variance in the weights and biases for the input sequence $X$ (Equation \ref{eq:wout}, Equation \ref{eq:bout}).
\begin{equation}
    W_{out} = XW_{\mu}^T + \sqrt{ (W_{\sigma}^2)^T X}*\epsilon
    \label{eq:wout}
\end{equation}
\begin{equation}
    b_{out} = Xb_{\mu}^T + b_\sigma * \epsilon
    \label{eq:bout}
\end{equation}
Then we calculate the KL divergence between our estimated posteriors and true priors for the weights and biases: $W_{KL} = KLD( 0, 1, W_\mu, W_{\sigma})$ and $b_{KL} = KLD( 0, 0.1, b_\mu, b_{\sigma})$. 
Finally, the forward computation sets a global KL divergence term ($KL=W_{KL} + b_{KL}$) and returns $W_{out} + b_{out}$.
The KL divergence term is used in the join loss function for tuning our variational posterior.

\section*{ 3.1.4 Encoder Feature Extraction }
Although the original Transformer architecture does not involve any convolutions, recent work in the Image Recognition domain (\shortcite{vgg-net}) has lent itself useful
for sequence-to-sequence ASR tasks \shortcite{joint-ctc-attn}. 
The modified VGG Network from \shortcite{joint-ctc-attn} is used in the Encoder to further enhance in the input feature set drawing ideas from unsupervised speech representation tasks
as seen in \shortcite{autoreg-speech-repr}, \shortcite{audio-classif-dbn}, and \shortcite{acoustic-modeling}. 
The output from this initial convolutional layer is passed to the encoder layers in the final network.

\section*{ 3.1.5 BayesSpeech Model }
\begin{figure}[!ht]
    \begin{subfigure}{0.5\textwidth}
        \centering
        \includegraphics[scale=0.15]{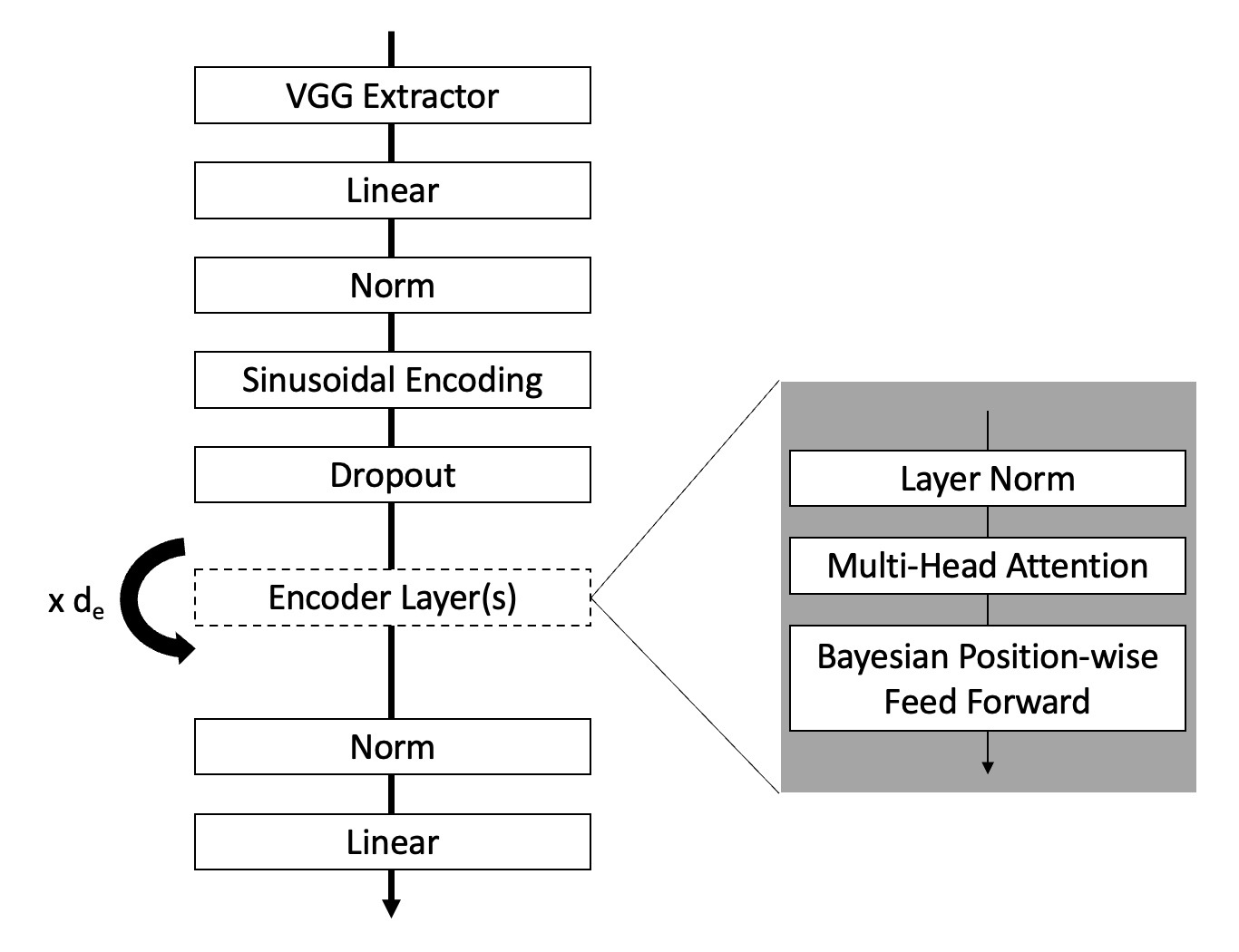}
        \caption{ BayesSpeech Encoder }
        \label{fig:modelsuba}
    \end{subfigure}
    \begin{subfigure}{0.5\textwidth}
        \centering
        \includegraphics[scale=0.15]{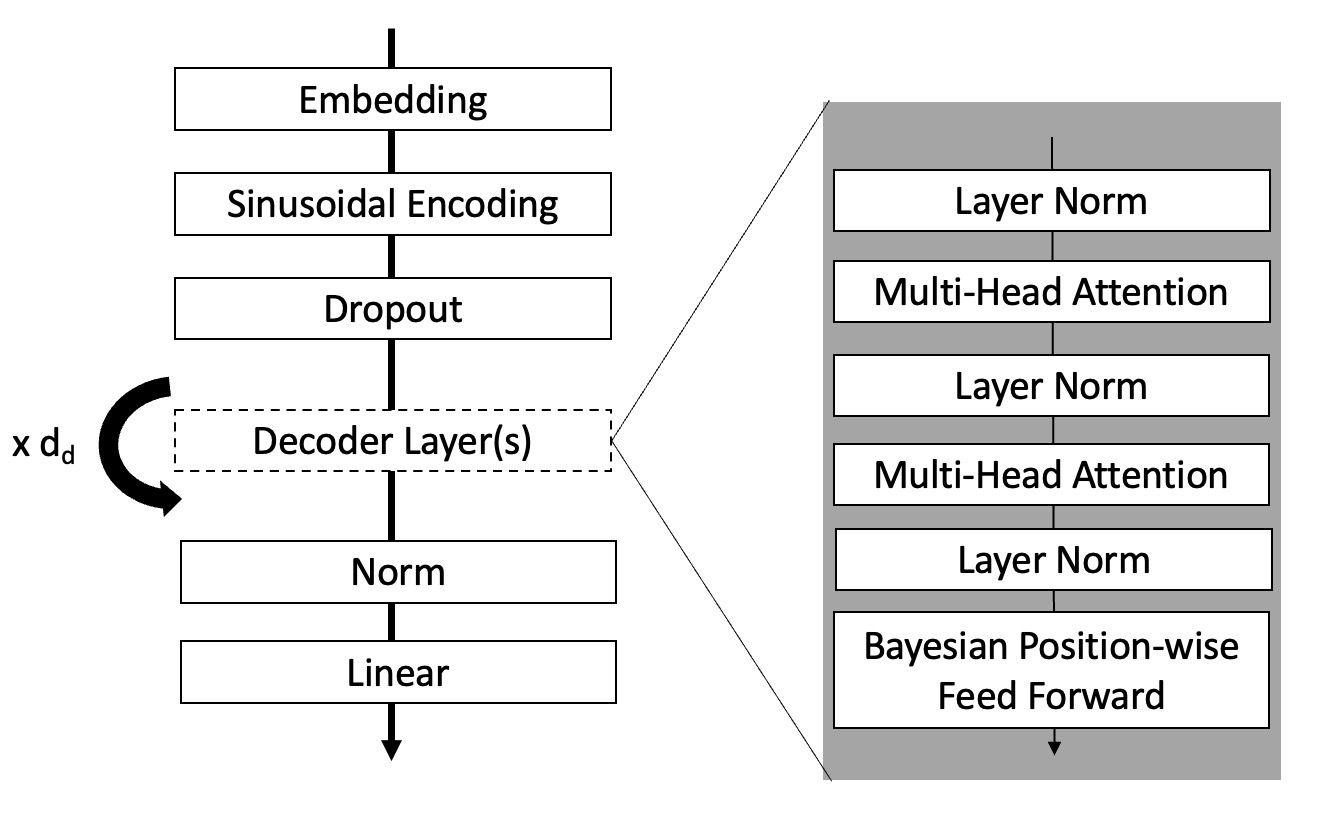}
        \caption{ BayesSpeech Decoder }
        \label{fig:modelsubb}
    \end{subfigure}
    \caption{ BayesSpeech Encoder and Decoder Diagrams }
    \label{fig:model}
\end{figure}

\vspace{-0.5em}
Putting this together, we arrive at our final model architecture (Figure \ref{fig:model}). 
The model passes the input through an Encoder (Figure \ref{fig:modelsuba}) and then passes the encoder output through a Decoder (Figure \ref{fig:modelsubb}).
In our model, we use 12 encoder block layers ($d_e = 12$) and 6 decoder block layers ($d_d = 6$).
These 18 inner layers contain a Mutli-Head attention block as well as a Bayesian Position-wise Feed Forward block.
The model has a dimension of 512 and a feed forward dimension of 2148.

\section*{ 3.2 Model Training}
\vspace{-0.5em}

In order to train our transformer model, we utilize a variation of the Bayes-By-Backprop algorithm \shortcite{weight-uncertainty} 
with a joint Connectionist Temporal Classification and Cross Entropy loss function (Joint CTC, CrossEntropy Loss).
The two-stage training is meant to: 
\begin{enumerate}
    \item further tune the sampling mechanics for the variational posterior distribution the weights are drawn from 
    \item and learn the temporal alignments and classification loss of the output tokens.
\end{enumerate}
We have found that trying to optimize each component separately leads to over-fitting in one of the domains of this problem.
If we choose a large step size and seek to minimize the aggregate KL divergence across the Bayesian layers, we cannot further learn the alignments.
And if we choose a small step-size and learn the alignments, we introduce too much randomness in the output for our results to be meaningful.
Therefore we introduce a scaling function, similar to the one in Bayes-by-Backprop for managing the tradeoff over epoch iterations (Minibatch Weighting).
Our model was trained on the LibriSpeech-960 dataset \shortcite{libri-speech}. 
The utterances in the dataset were converted to Mel Spectrogram form with 80 channels a width of 20ms and a stride of 10ms.

\section*{ 3.2.1 Tuning Variational Posterior }
\vspace{-0.5em}

The Bayesian part of our model tries to fit a variational posterior distribution ($q_\theta$) to a true intractable posterior ($p$) for each of the weights in the network.
In order to do so, we reduce the problem of fitting $q_\theta$ to that of a Minimum Description Length (MDL) problem \shortcite{min-desc-len-weights} \shortcite{modeling-by-sdd} \shortcite{keeping-nn-simple}.
While we have introduced the explicit calculation of the Kullback-Leibler divergence between our variational posterior and true prior above ($D_{KL}( q_\theta || p )$),
it is important to conceptualize the divergence as the MDL problem.

The Minimum Description Length principal is that the best model for a given dataset balances the tradeoff between describing the model and
describing the misfit between the model and the data \shortcite{keeping-nn-simple}.
The KL divergence criteria we use has the goal of keeping weights simple by penalizing the amount of information they contain.
Ultimately, this methodology will lead to a better separation between prediction accuracy and model complexity and is explicitly differentiable \shortcite{practical-vi}.
The variational loss function used has two parts:
\begin{enumerate}
    \item Error Loss - the expected value of negative log probability in samples from $q_\theta(\beta)$ (where $\beta$ are the model's parameters)
    \item Complexity Loss - the KL divergence between the tractable, variational posterior and the parameterized prior, $D_{KL}(q_\theta(\beta) || p_\alpha)$.
\end{enumerate}

\begin{figure}[!ht]
    \begin{subfigure}{0.5\textwidth}
        \centering
        \includegraphics[scale=0.35]{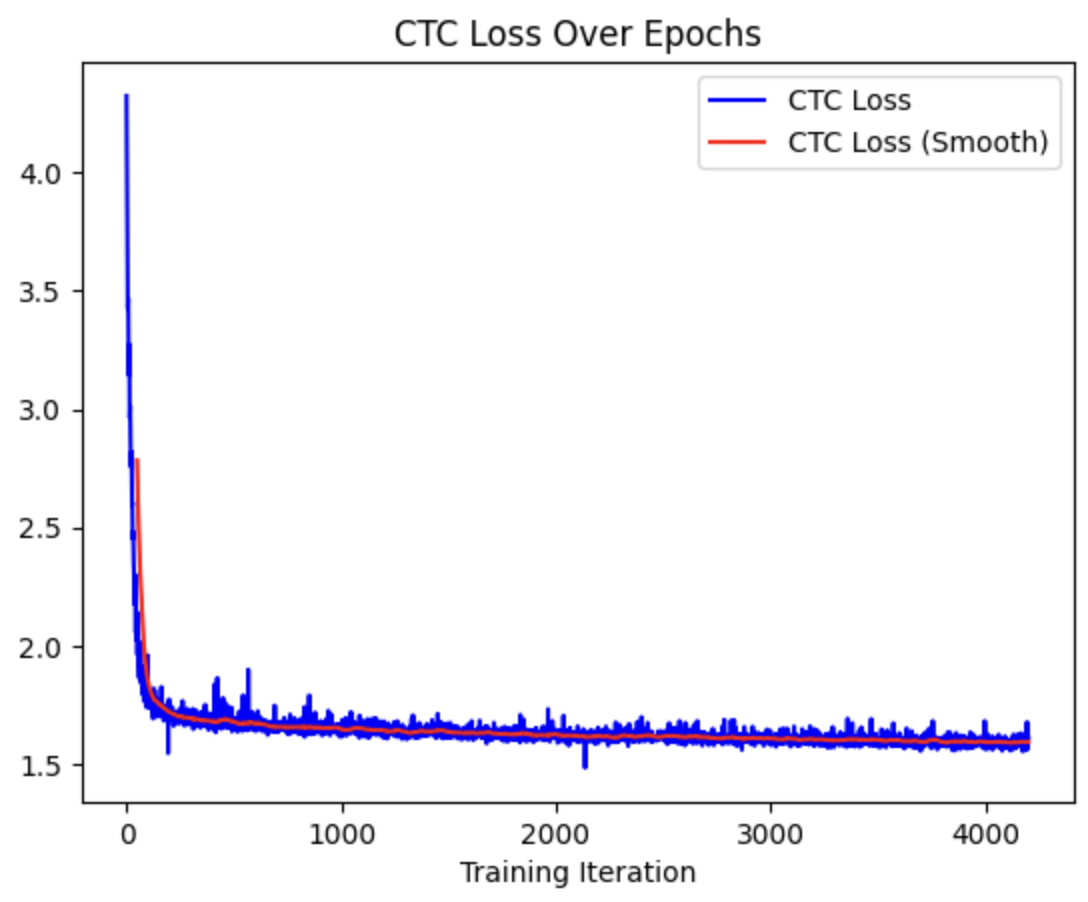}
        \caption{ CTC Loss Over Training Iterations }
        \label{fig:trainingsuba}
    \end{subfigure}
    \begin{subfigure}{0.5\textwidth}
        \centering
        \includegraphics[scale=0.35]{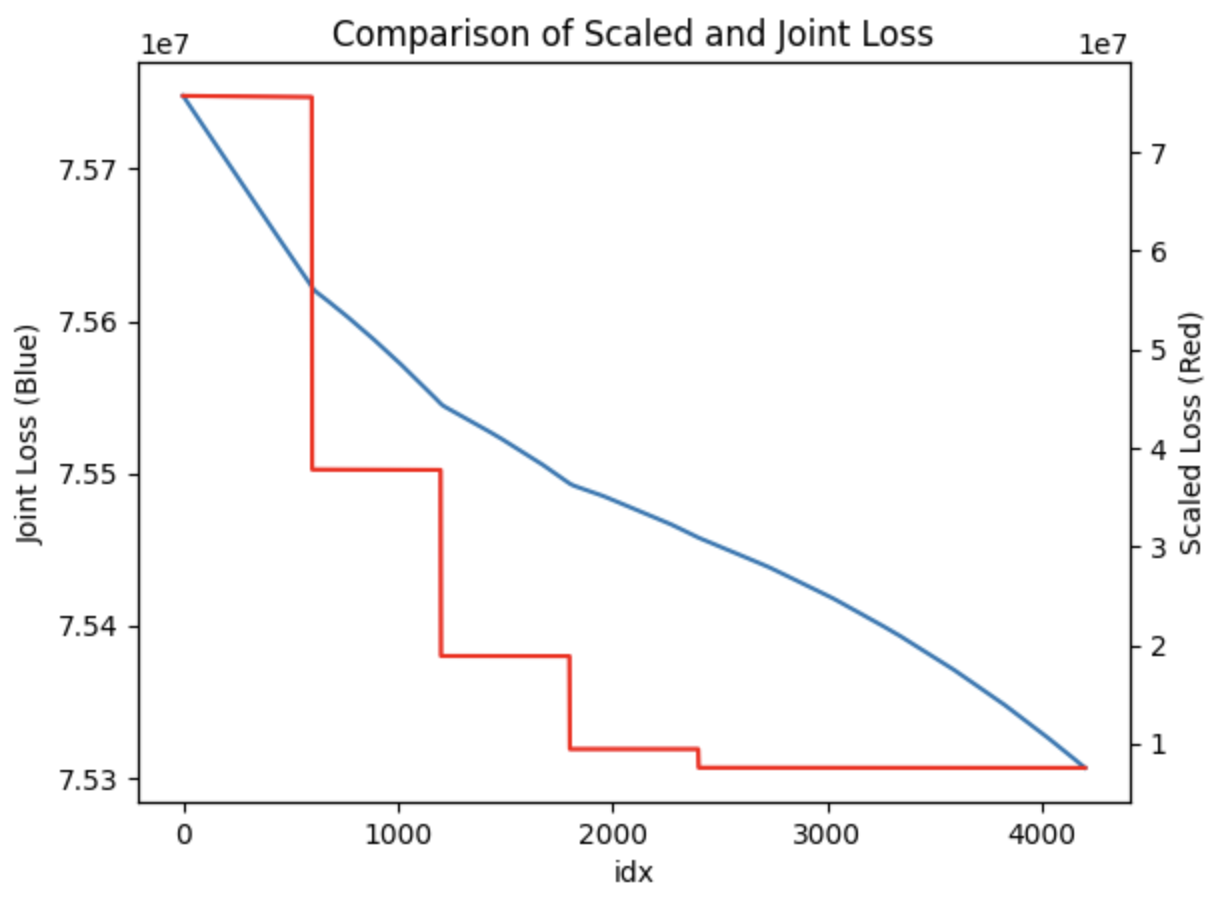}
        \caption{ Joint (Blue) and Scaled (Red) Loss Over Time }
        \label{fig:trainingsubb}
    \end{subfigure}
    \caption{ Loss Functions over Training Iterations }
    \label{fig:training}
\end{figure}

In each batch, we seek to gently tune our model's variational posterior ($q_\theta$) to continue random sampling but isolate different
weights that have different levels of kurtosis.
Due to the minibatch weighting, discussed in a later section, we see a consistent decline in the joint loss value dominated by the KL divergence term (blue, Figure \ref{fig:trainingsubb}).
Blundell et al. also show that using this relative kurtosis can create thinner models with an explicit scheme for weight pruning.
Weights that are more leptokurtic are kept while platykurtic ones are discarded.
While this is beyond the scope of this paper, it would present and interesting future research case for the model presented.

\section*{ 3.2.2 Joint CTC, CrossEntropy Loss}
\vspace{-0.5em}

In order to penalize the model for alignment of the input sequence to the output tokens, we utilize a joint Connectionist Transporal Classification \shortcite{CTC-graves} and Cross Entropy Loss function.
The goal of this two term loss function is to manage a gradient through the alignment of tokens in the feature input (CTC) as well as the actual classification loss of the aligned output and the true tokens.
The loss functions weights the two as so: $L(X) = 0.3 * CTC(X) + 0.7 * CE(X)$. 
We do this in order to help smooth out the gradient while maintaining the proper loss to back-propagate through the network.
Due to the adversarial nature of the Bayesian outputs, we find that this joint loss descends rapidly then continues to descend without adjustment to the original learning rate (Figure \ref{fig:trainingsuba}).
In our training, we held the learning rate fixed at $10^{-6}$.

\section*{ 3.2.3 Minibatch Weighting }
\vspace{-0.5em}
Blundell et al. found that earlier epochs have a greater importance on tuning of the variational posterior than later ones.
We adopt a similar methodology where we weight the KL divergence term according to the epoch ($e$) and number of epochs ($n_e$) (Equation). 
\begin{equation}
    MinibatchWeight( e, n_e ) = \frac{2^{n_e - e}}{2^{n_e} - e}
    \label{eq:minibatch}
\end{equation}
To better aid training over time, we choose an epoch indexer where the epoch index is integer divided by 10.
When the training loop runs for multiple hours, this helps keep the KL divergence more heavily weighted at first.
We then weight the KL divergence term by the minibatch weight term (Equation \ref{eq:fulltraining}).
The $KL_{div}$ term is the sum of all KL divergences over the Bayesian layers.
\begin{equation}
    L( X, e, n_e ) = MinibatchWeight( e, n_e ) * KL_{div} + 0.3 * CTC(X) + 0.7 * CE(X) 
    \label{eq:fulltraining}
\end{equation}

\section*{ 4. Results }
\vspace{-0.5em}

We split our model into two variants: one that outputs a character sequence and one that outputs tokenized word-pieces from a Sentencepiece language model with vocab size of 1000 \shortcite{sentencepiece}.
We trained each model variant on a single A-100 GPU through Google Colab for 8 hours with a batchsize of 24.

\begin{table}[!ht]
    \center
    \begin{tabular}{lcccc}
    \multicolumn{1}{c}{{\ul Model}} & \multicolumn{2}{c}{{\ul WER (w/o LM)}} & \multicolumn{2}{c}{{\ul WER (w/ LM)}} \\
    \multicolumn{1}{c}{}            & {\ul test-clean}   & {\ul test-other}  & {\ul test-clean}  & {\ul test-other}  \\
    LAS \shortcite{LAS}                            & 2.89\%             & 6.98\%            & 2.33\%            & 5.17\%            \\
    Transformer \shortcite{attn-is-all-u-need} & 2.4\%              & 5.6\%             & 2.0\%             & 4.6\%             \\
    Conformer \shortcite{conformer}                      & 2.1\%              & 5.0\%             & 2.0\%             & 4.3\%             \\
    \textbf{BayesSpeech}            & 4.5\%              & 6.5\%             & 4.0\%             & 5.7\%            
    \end{tabular}
    \caption{ WER Results on LibriSpeech dataset \label{table:results}}
\end{table}

As shown in Table 1, our model performs nearly as well as the state of the art ASR models.
Our Bayes speech model reaches respectable Word Error Rates with and without a language model on the LibriSpeech dataset.
The Bayes Model as well was trained for just 8 hours on a single GPU.
For instance, the Conformer model was trained over the course of multiple days on multiple GPUs (8).
During evaluation, we use beam search with a beam width of 10 over the set of possible decoded sequences.
This appears to be the standard decoding methodology giving the probabilistic output of the model's decoder.

When the input sequence passes through our Bayesian feed forward layers, we believe this creates an adversarial input stream.
Rather than artificially augment the input Mel Spectrogram inputs \shortcite{spec-augment}, these layers produce a probabilistic
feature encoding of the input.
We believe that this general adversarial training technique allows our model to converge faster with less training time and resources.
The randomness introduced in the model also helps better contextualize outputs.
As we continue to tune the variational posterior over the weights, I imagine we would see a dramatic increase in performance.
Because our model yielded reasonable results after 8 hours, we stopped training but future work may investigate if increased training could
further improve our performance.
There may also be a benefit to equally weighting the variational component and the CTC loss component of the global loss function.
Similarly, in future work it may be useful to explore systematic model pruning as presented in Blundell et al.

\section*{ 5. Conclusion}
\vspace{-0.5em}
Currently, best in class Automatic Speech Recognition solutions require multiple days of training on multiple GPUs.
These models also take a frequentist approach to weight training. 
In this work, we present BayesSpeech; a Bayesian Transformer Network for learning an intractable posterior distribution over which weights are drawn
in feed forward layers.
We believe this probabilistic encoding of the input feature set creates a better representation of the input Mel Spectrogram.
This mechanic in conjunction with a joint loss function yields near state-of-the-art results on the LibriSpeech dataset.

\newpage
\bibliographystyle{apacite}
\bibliography{refs.bib}

\end{document}